\documentstyle[12pt]{article}

\begin{document}
\title{ Statistical properties of Klauder-Perelomov coherent
 states for the Morse potential}
\author{{\bf M. Daoud}$^{1,}${\footnote{Permanent adress: LPMC, Faculty of Sciences, University Ibn Zohr, Agadir, Morocco}}  and  {\bf D. Popov}$^2$ \\
\\
$^1$Max Planck Institute for the Physics of Complex Systems\\
Dresden, Germany\
\\
$^2$University "Politehnica" of Timi\c soara,  Department of Physics\\
Timi\c soara, Romania}

\maketitle

\begin{abstract}
We present in this letter a realistic construction of the coherent
states for the Morse potential using the Klauder-Perelomov
approach . We discuss the statistical properties of these states,
by deducing the Q- and P-distribution functions. The thermal
expectations for the quantum canonical ideal gas of the Morse
oscillators are also calculated.
\end{abstract}

\vfill
\newpage 

\section{Introduction}

The one-dimensional Morse potential was first introduced as a
useful model for diatomic molecules in 1929 \cite{morse}. Since
its introduction, the Morse oscillator has proved very useful for
various problems in diverse fields of physics and chemistry
(diatomic and polyatomic molecular systems, quantum chemistry,
spectroscopy, chemical bonds) \cite{pauling, nieto, sage, vasan,
dahl, popov-thesis} (and the references therein).

On the other hand, since the pioneer papers of Glauber, Klauder
and Perelomov about the coherent states of the one-dimensional
harmonic oscillator \cite{kla-ska}, \cite{perelomov}, and their
different applications in physics \cite{Ali}, this approach has
been paid attention also for other potentials. The three way to
define the coherent states for the one-dimensional harmonic
oscillator (i.e.: 1. as states which minimize the uncertainty
relations; 2. as states which diagonalize the lowering operator
and 3. as states obtained by action of the displacement operator
on the bound state), which in this case are convergent (lead to
the same results), for an another anharmonic potential lead to the
three kind of states, generically named: Barut-Girardello,
Klauder-Perelomov and Gazeau-Klauder \cite{klauder},
\cite{gazeau}, \cite{antoine}.

Based on these papers, the coherent states was constructed for a
series of potentials (see, \cite{daoud}, \cite{kinani-JPA34},
\cite{kinani-PLA}, \cite{kinani-IJMP}, \cite{kinani-JPA35}).

For the pseudoharmonic oscillator the Barut-Girardello coherent
states , as well as the photon-added Barut-Girardello coherent
states was constructed \cite{popov-JPA2001}, \cite{popov-JPA2002}.(see also, \cite{daoud-JPA2002}).

For the Morse oscillator, the construction of generalized coherent
states was begin by the papers of \cite{nieto}. Recently, Dong
have constructed the coherent states based on the SU(2)
realization for the Morse potential \cite{dong}, Fakhri and
Chenaghlou have constructed the Barut-Girardello coherent states
for the Morse potential \cite{fakhri}, while Roy and Roy , and,
also, Popov  have constructed and examined the properties of the
Gazeau-Klauder coherent states for the same potential
\cite{roy-roy}, \cite{popov-PLA}. So, the examination of the
Klauder-Perelomov coherent states for the Morse potential, which is
the main aim of this paper becomes a naturally task.

This paper is organized as follows. In Sec. 2, after a brief
recall of the main facts and properties of the Morse oscillator,
we deduce the corresponding raising and lowering operators and we
construct the Klauder-Perelomov coherent states for this
oscillator. Section 3 is devoted to the examination of the
statistical properties of these coherent states, by examining a
quantum canonical ideal gas of Morse oscillators, in
thermodynamical equilibrium with the reservoir. the concluding
remarks will be given in Sec. 4.

\section{Klauder-Perelomov coherent states}

We shall begin by recalling some basic facts of the Morse
potential, mainely the eigenstates and corresponding creation and
annihilation operators. The hamiltonian for a one dimensional
quantum particle evolving in the Morse potential is given by

\begin{equation}\label{H-l}
H^l _- = -\frac{d^2}{d^2 x} + V^l (x)
\end{equation}
where

\begin{equation}\label{V-l}
V^l(x) = (l+1)^2 - (2l+3)e^{-x} + e^{-2x}
\end{equation}
with $l$ is an integer to simplify. The energies are given by
\begin{equation}\label{E-n-l}
E^{l}_n = (l+1)^2 - (l+1-n)^2
\end{equation}
where the quantum number $n$ takes the finite sequence of values: $0, 1, ..., l$.

The discrete spectrum of the Morse potential is finite and this
fact is very important in the construction of his coherent states
as we will discuss in the sequel of this note.

Because the ground state is zero energy, it is well known that one
can factorize the Hamiltonian as
\begin{equation}\label{H-l-minus}
H^{l}_{-} = X^{+}_{l}X^{-}_{l}
\end{equation}
where the operators $X^{\pm}_{l}$ (hermitian conjugated of each other,i.e
$(X^{+}_{l})^{\dagger}
= X^{-}_{l}$) are given by
\begin{equation}\label{X-pm-l}
X^{\pm}_{l} = \mp \frac{d}{dx} - (e^{-x} - (l+1)).
\end{equation}

From the supersymmetric quantum mechanics, it is well established
that the supersymmetric partners
\begin{equation}\label{H-l-mp}
 H^{l}_{\mp}= X^{\pm}_{l}X^{\mp}_{l} = - \frac{d^2}{dx^2} + e^{-2x} - (2(l+1) \pm 1)e^{-x} + (l+1)^2
\end{equation}
have the same energy spectra, but different eigenstates. Indeed,
it is easy to verify that
\begin{equation}\label{H-H}
 H^{l}_{+} = H^{l-1}_{-} + (2l+1)
\end{equation}
which implies that the eigenstates of $H^{l}_{+}$ can be obtained from
$ \vert \psi_n^{l}\rangle $ corresponding to $H^{l}_{-}$ by substituting
$l$ by $l-1$. The Schr\"odinger equations for $H^{l}_{\mp}$ are

\begin{equation}\label{H-mp-psi}
H^{l}_{\mp}\vert \psi_n^{l-\frac{1}{2}\pm s}\rangle = (l+1)^2 -
(l+ \frac{1}{2} \pm s - n )^2 \vert \psi_n^{l-\frac{1}{2}\pm
s}\rangle ,
\end{equation}
where $ s = \frac{1}{2}$. It is straightforward to check, for each
$l$, the intervening relations
\begin{equation}\label{X-H-H-X}
X^{\mp}_{l}H^{l}_{\mp}= H^{l}_{\pm}X^{\mp}_{l}.
\end{equation}

Due to the latter relations the operators $X^{\pm}_{l}$ link the
Hilbert spaces ${\cal H}^{l\pm s -\frac{1}{2}}$ and ${\cal
H}^{l\mp s - \frac{1}{2}}$ spanned by the bounded states of the
Morse potential. They connect eigenfunctions with the same energy.
Indeed one has

\begin{equation}\label{X-minus-psi}
X_{-}^{l} \vert \psi^{l}_{n}\rangle = \sqrt {E^{l}_n} e^{i\alpha
(2(l-n)+1)} \vert \psi_{n-1}^{l-1} \rangle
\end{equation}
\begin{equation}\label{X-plus-psi}
X_{+}^{l} \vert \psi^{l-1}_{n}\rangle = \sqrt {E^{l}_{n+1}}
e^{-i\alpha (2(l-n)+1)} \vert \psi_{n+1}^{l} \rangle
\end{equation}
where $\alpha \in {\bf R}$. The operators $X_{-}^{l} $ and $X_{+}^{l}$
 are not
the ladder operators of the Hamiltonian $H^{l}_{-}$. To define the creation and
annihilation operators for $H^{l}_{-}$, we define the operator
\begin{equation}\label{U}
U = \sum_{m }  \vert \psi_{m}^{l} \rangle \langle \psi_{m}^{l-1}
\vert
\end{equation}
satisfying
$U^+ U = 1$ and $\vert  \psi_{n-1}^{l} \rangle = U \vert  \psi_{n-1}^{l-1} \rangle$.
Using the transformation $U$, define a new pair of operators
\begin{equation}\label{A-X-U}
A_{+}^{l} = X_{+}^{l} U^+ {\hskip 2cm},{\hskip 2cm} A_{-}^{l} = U
X_{-}^{l}
\end{equation}

Combining (\ref{X-minus-psi}), (\ref{X-plus-psi}) and (\ref{U}),
we get the actions of $A_{+}^{l}$ and $A_{-}^{l}$ on the
eigenstates $\{ \vert \psi_n^l\rangle \}$ as
\begin{equation}\label{A-minus}
A_{-}^{l} \vert \psi_{n}^l \rangle = \sqrt {E_{n}^l} e^{i\alpha
(2(l-n)+1)} \vert \psi_{n-1}^l \rangle
\end{equation}
\begin{equation}\label{A-plus}
A_{+}^{l}\vert \psi_{n} \rangle = \sqrt {E_{n+1}^l} e^{-i\alpha
(2(l-n)+1)} \vert \psi_{n+1}^l \rangle
\end{equation}

The set of operators $\{ A_{-}^{l}, A_{+}^{l}\}$ are the creation
and annihilation operators of $H^l = A_{+}^{l}A_{-}^{l} =
X_{+}^{l}X_{-}^{l}$. They satisfy the following commutation
relations
\begin{equation}\label{comut}
[ A_{-}^{l} , A_{+}^{l} ] = - 2 N + (2l + 1)
\end{equation}
where the operator $N$ is given by
\begin{equation}\label{N-oper}
N \vert \psi_{n}^l \rangle  = n  \vert \psi_{n}^l \rangle.
\end{equation}

The algebra generated by $A_{-}^{l}$ , $A_{+}^{l}$ and $N$
provides the appropriate tool to build up the coherent states for
the Morse potential.

The so-called Klauder-Perelomov coherent states for quantum
system embedded in the Morse potential are defined
\begin{equation}\label{CS-def}
\vert z , \alpha \rangle = \exp ( z A_{+}^{l} - \bar z A_{-}^{l})
\vert \psi_0^l \rangle
\end{equation}

Expanding the displacement operator and using the actions of the
creation and annihilation operators, it comes
\begin{equation}\label{CS-intermed}
\vert z , \alpha \rangle = \sum_{n=0}^l z^nI_n^l(|z|)e^{-i \alpha
E_n}\vert\psi_n^l\rangle.
\end{equation}
where
\begin{equation}\label{I-n-l}
I_n^l(|z|) = \sum_{j=0}^{\infty} \frac{(-|z|^2)^j}{(n+2j)!}
\Delta^l(n+1,j)
\end{equation}

The quantities $\Delta$ which occur in the last formula are give
by
\begin{equation}\label{Delta}
\Delta^l(n+1,j) =
\frac{n!(2l+1)!}{(2l+1-n)!}\sum_{i_{1}=1}^{n+1}E^l_{i_1}\sum_{i_{2}=1}^{i_{1}+1}E^l_{i_2}...\sum_{i_j=1}^{i_{j-1}+1}E^l_{i_j}
\end{equation}
with $\Delta^l(n+1,0) = \frac{n!(2l+1)!}{(2l+1-n)!}$. They satisfy the following reccurence equation
\begin{equation}\label{Delta-Delta}
\Delta^l(n+1,j) = \sqrt{2n(l+1)-n^2} \Delta^l(n,j) +
\sqrt{2nl+2l+1-n^2} \Delta^l(n+2,j-1).
\end{equation}

Setting
\begin{equation}\label{J-n-l}
J^l_n(|z|) = |z|^n\sqrt{\frac{(2l+1-n)!}{n!(2l+1)!}} I_n^l(|z|),
\end{equation}
we get the first order differential equation
\begin{equation}\label{deriv-J}
\frac{dJ^l_n(|z|)}{d|z|} = J^l_{n-1}(|z|) - (2nl+2l+1-n^2)
J^l_{n+1}(|z|).
\end{equation}

The solution of this equation takes the simple form
\begin{equation}\label{J-cos}
J^l_n(|z|) = \frac{1}{n!} (\cos (|z|))^{l-1}(\ tg (|z|))^n,
\end{equation}
and the Morse coherent states rewrite as
\begin{equation}\label{CS-final}
\vert Z , \alpha \rangle = (1 + |Z|^2)^{-\frac{l}{2}}
\sum_{n=0}^{l} \sqrt{\frac{l!}{n! (l-n)!}} Z^n e^{-i\alpha
E_{n}^l}\vert \psi_{n}^l \rangle
\end{equation}
where $Z = \frac{z}{|z|} \ tg (|z|)$. They have the property of
strong continuity in the label space and completeness in the sense
that there exists a positive measure such that they solve the
resolution to identity. The appropriate form of this resolution is
\begin{equation}\label{resolution}
\int d\mu (Z , \bar Z) \vert Z, \alpha \rangle \langle Z , \alpha
\vert = \sum_{n=0}^{l} \vert \psi_{n}^l\rangle \langle \psi_{n}^l
\vert.
\end{equation}

To determine the measure $d\mu (Z , \bar Z)$, we assume his isotropy and we set
\begin{equation}\label{measure}
d\mu(Z,\bar Z) = (1 + \vert Z \vert^2)^{l} h( \vert Z \vert ^2)
\vert Z \vert d \vert Z \vert d\theta /\pi
\end{equation}
with $Z = \vert Z \vert e^{i\theta}$. Substituting Eq.
(\ref{measure}) into Eq.(\ref{resolution}), we obtain the
following sum which should be satisfied by the function $h(x =
\vert Z \vert ^2 ))$
\begin{equation}\label{int-h}
\int_{0}^{\infty} x^n h(x) dx = \frac{n!(l-n)!}{l!}.
\end{equation}

The inverse Mellin transform gives
\begin{equation}\label{h-final}
h(x) = \frac{l+1}{(1+x)^{l+2}}.
\end{equation}

This result can be obtained also by using the definition of
Meijer's G-function and the Mellin inversion theorem
\cite{math-sax}.

Due to phase introduced in the definitions of annihilation and
creation operators, the obtained coherent states are temporally
stable. Indeed
\begin{equation}\label{temporally}
e^{-it H_{-}^l} \vert Z, \alpha \rangle = \vert Z, \alpha + t
\rangle.
\end{equation}

The physical utility of Klauder-Perelomov coherent sates for the
Morse potential (\ref{CS-final}) in different applications
consists in the calculation of the expectation (mean) values of a
certain physical observable $A$ which characterizes the quantum
system embedded in the Morse potential, with respect to
$|Z,\alpha>$:

\begin{eqnarray}\label{expect-value}
<Z,\alpha|A|Z,\alpha>\equiv
    <A>_{Z}= \qquad \qquad \qquad \qquad \\
    =(1+|Z|^2)^{-l}\sum_{n,m=0}^{l}\sqrt{{l\choose n}{l\choose
    m}}Z^{n}\overline{Z}^{m}e^{-\mathrm{i}\alpha(E_{n}^{l}-E_{m}^{l})}<\psi_{m}^{l}|A|\psi_{n}^{l}>
\end{eqnarray}
where we have used the binomial coefficients \cite{gradshteyn}:

\begin{equation}\label{binomial}
{l\choose n}=\frac{l!}{n!
    (l-n)!}=\frac{\Gamma(l+1)}{\Gamma(n+1)\Gamma(l+1-n)}.
\end{equation}

From a practical point of view, the most important operators are
the diagonal operators in the eigenstates-basis $|\psi_{n}^{l}>$
corresponding to $H_{-}^{l}$, so that

\begin{equation}\label{A-diag}
<\psi_{m}^{l}|A|\psi_{n}^{l}>=a_{n}\delta_{mn}.
\end{equation}

If $A=N^{s}$, where the operator $N$ is given by Eq.
(\ref{N-oper}) and $s$ is an integer, we have:

\begin{equation}\label{N-s-diag}
<A>_{Z}=\frac{1}{(1+x)^{l}}\sum_{n=0}^{l}{l\choose
    n}n^{s}x^{n}=\frac{1}{(1+x)^{l}}\left(x\frac{d}{dx}\right)^{s}(1+x)^{l},
\end{equation}
where we have used the simplified notation: $x=|Z|^{2}$.

For $s=1$ and $2$ we obtain, successively:

\begin{equation}\label{N-1-z}
<N>_{Z}=l\frac{x}{1+x},
\end{equation}

\begin{equation}\label{N-2-z}
<N^{2}>_{Z}=l\frac{x}{1+x}+l(l-1)\left(\frac{x}{1+x}\right)^{2}.
\end{equation}

These expectation values are useful in order to calculate the
second-order correlation function for the morse oscillator
\cite{walls}:

\begin{equation}\label{g-2-z}
(g^{2})_{Z}=\frac{<N^{2}>_{Z}-<N>_{Z}}{(<N>_{Z})^{2}}=\frac{l-1}{l}<1.
\end{equation}

Moreover, the Mandel Q-parameter is \cite{solomon}

\begin{equation}\label{Q-z}
Q_{Z}=<N>_{Z}\left[(g^{2})_{Z}-1\right]=-\frac{x}{1+x}<0.
\end{equation}

These quantities provide information about the inherent
statistical properties of the Klauder-Perelomov coherent states of
the Morse oscillator $|Z, \alpha>$. These properties depend on the
analytical expressions of the functions (\ref{g-2-z}) and
(\ref{Q-z}) as depending on the variable $x=|Z|^{2}$. Because of
the structure of these functions, $(g^{2})_{Z}$ and $Q_{Z}$ can be
evaluated analytically. Generally speaking, the states $|Z,
\alpha>$ exhibit sub-Poissonian statistics for those values of
$x=|Z|^{2}$ for which $Q_{Z}<0$ (or $(g^{2})_{Z}<1$)(antibunching
effect), Poisson statistics for values for which $Q_{Z}=0$ (or
$(g^{2})_{Z}=1$) and supra-Poissonian statistics for values of $Z$
for which $Q_{Z}>0$ (or $(g^{2})_{Z}>1$) (bunching effect). As we
see from the above equations, the Klauder-Perelomov coherent
states of the Morse oscillator $|Z, \alpha>$ are sub-Poissonian,
for all values of the variable $|Z|^{2}$.

By putting $A=H_{-}^l$ and using the eigenvalue equation
(\ref{E-n-l}), we obtain the action identity:

\begin{eqnarray}\label{action-identity}
<H_{-}^l>_{Z}=\frac{1}{(1+x)^{l}}\sum_{n=0}^{l}{l\choose
    n}x^{n}E_{n}^{l}=\qquad \qquad \qquad  \\ \nonumber
    =2(l+1)<N>_{Z}-<N^{2}>_{Z}=l(2l+1)\frac{x}{1+x}-l(l-1)\left(\frac{x}{1+x}\right)^{2}\equiv
    f(x).
\end{eqnarray}

We see that this is a certain function of the label $x=|Z|^{2}$
 and this fact characterizes all the coherent states
corresponding to systems with a finite dimensional energy
spectrum, as it was mentioned earlier \cite{roy-roy},
\cite{popov-PLA}.

\section{\bf Statistical properties}

We consider a quantum gas of the Morse oscillators in
thermodynamic equilibrium with the reservoir (the thermostat) at
temperature $T$, which obeys the quantum canonical distribution.
Also, we consider that the individual Morse oscillators are in
such states which are labelled by number state vectors
$|\psi_{n}^{l}>$. The corresponding normalized density operator
for a fixed $l$-parameter (or, equivalently, for a fixed Morse
oscillator) is then

\begin{equation}\label{rho-l}
\rho_{l}=\frac{1}{Z_{l}} \sum_{n=0}^{l}e^{-\beta
\varepsilon_{n}^{l}} |\psi_{n}^{l}><\psi_{n}^{l}|,
\end{equation}
where $\beta = (k_{B}T)^{-1}$, $k_{B}$ is Boltzmann's constant and
$Z_{l}$ - the normalization constant, i.e. the partition function
for a certain fixed parameter $l$:

\begin{equation}\label{partition}
Z_{l}=\sum_{n=0}^{l}e^{-\beta
\varepsilon_{n}^{l}}
\end{equation}

In the above exponential it appear the dimensional Morse
eigenenergies:

\begin{equation}\label{E-dim}
 \varepsilon_{n}^{l}\equiv \frac{\hbar
    \omega}{2(l+1)}E_{n}^{l}=\hbar \omega n-\frac{\hbar
    \omega}{2(l+1)} n^{2}
\end{equation}
where $\omega$ is the angular frequency for the Morse oscillator
with a fixed parameter $l$.

The Q-function (the Husimi's function), i.e. the diagonal elements
of the density operator in the representation of coherent states,
is

\begin{equation}\label{rho-z}
<Z, \alpha|\rho_{l}|Z, \alpha>=\frac{1}{Z_{l}}\sum_{n=0}^{l}
e^{-\beta \varepsilon_{n}^{l}}{l\choose n} x^{n}.
\end{equation}

It is not difficult to verify that the normalization condition of
the density operator is accomplished:

\begin{equation}\label{trace-rho}
Tr \rho_{l}=\int d\mu(Z,\overline{Z})<Z, \alpha|\rho_{l}|Z,
    \alpha>=1
\end{equation}
where we have used the following integral \cite{gradshteyn},
\cite{proudnikov}:

\begin{equation}\label{int-1}
\int_{0}^{\infty}dx x^{\alpha
    -1}(x+z)^{-\gamma}=z^{\alpha-\gamma}B(\alpha, \gamma-\alpha)=
    z^{\alpha-\gamma}\frac{\Gamma(\alpha)\Gamma(\gamma-\alpha)}{\Gamma(\gamma)}.
\end{equation}

If we consider the dimensional Morse Hamiltonian, and the
dimensional eigenvalues of energy, instead of $H_{-}^{l}$ and
$E_{n}^{l}$, then the corresponding energy exponential can be
written as follows:

\begin{equation}\label{beta-E}
\beta \varepsilon_{n}^{l}=\beta \hbar\omega
    n-\beta\frac{\hbar\omega}{2(l+1)}n^{2}\equiv
    {\mathcal{A}}n-{\mathcal{B}}n^{2}.
\end{equation}

For most of the diatomic molecules $\mathcal{B}\ll \mathcal{A}$.
So, the limits of the parameter $2(l+1)$ are very large, e.g.
37.1586 for $\mathrm{H}_{2}$ molecule, i.e. for a "light" molecule
and 348.78 for $\mathrm{I}_{2}$, a "heavy" molecule
\cite{popov-ps}. As a consequence, the quantity $\mathcal{B}$ can
be regarded as a perturbation constant and the energy exponential
can be expanded in the power series as follows:

\begin{equation}\label{expand-E}
e^{-\beta \varepsilon_{n}^{l}} =
e^{-{\mathcal{A}}n}\sum_{k=0}^{\infty}
\frac{{{\mathcal{B}}}^{k}}{k!}n^{2k}=\sum_{k=0}^{\infty}\frac{{{\mathcal{B}}}^{k}}{k!}
\left(\frac{d}{d{\mathcal{A}}}\right)^{2k}\left[e^{-{\mathcal{A}}}\right]^{n}.
\end{equation}

We can also use  the following operator identity:

\begin{equation}\label{op-ident}
\sum_{k=0}^{\infty}\frac{{{\mathcal{B}}}^{k}}{k!}\left(\frac{d}{d{\mathcal{A}}}\right)^{2k}\equiv
\exp{\left[{\mathcal{B}}\left(\frac{d}{d{\mathcal{A}}}\right)^{2}\right]}.
\end{equation}

Taking into account this ansatz, after the straightforward
calculations, we can write the Q-function in the following manner:

\begin{equation}\label{Q-function-final}
 <Z, \alpha|\rho_{l}|Z, \alpha>=\frac{1}{Z_{l}}\exp{\left[{\mathcal{B}}\left(\frac{d}{d{\mathcal{A}}}\right)^{2}\right]}
    \left(\frac{1+xe^{-{\mathcal{A}}}}{1+x}\right)^{l}.
\end{equation}

Because the Klauder-Perelomov coherent states for the Morse
oscillator $|Z,\alpha>$ form an overcomplete set of states, they
may be used as a basis set despite the fact that they are not
orthogonal. Let us perform the diagonal expansion of the density
operator $\rho_{l}$ in the coherent states basis:

\begin{equation}\label{rho-diagonal}
\rho_{l}=\frac{1}{Z_{l}}\int d\mu(Z,\overline{Z})
|Z,\alpha>P_{l}(|Z|^{2})<Z,\alpha|.
\end{equation}

In order to find the quasi-probability distribution function
$P_{l}(|Z|^{2})$ (or the P-function) from the above diagonal
expansion, we observe that the equation

\begin{equation}\label{f-rho-g}
<f|\rho_{l}|g>=\frac{1}{Z_{l}}\int d\mu(Z,\overline{Z})<f|Z,
\alpha>P_{l}(|Z|^{2})<Z, \alpha|g>
\end{equation}
must be fulfilled for any arbitrary vectors $<f|$ and $|g>$ from
the Hilbert space (or, for any vectors from the basis $|Z,
\alpha>$ or $|\psi_{n}^{l}>$).

The left-hand side of this equation is

\begin{equation}\label{LHS}
LHS=\frac{1}{Z_{l}} \sum_{n=0}^{l}e^{-\beta \varepsilon_{n}^{l}}
<f|\psi_{n}^{l}><\psi_{n}^{l}|g>,
\end{equation}
while, after the angular integration

\begin{equation}\label{angular}
\frac{1}{2\pi}\int_{-\pi}^{\pi}d\theta
    e^{-{\mathrm{i}}(n-m)\theta}=\delta_{nm},
\end{equation}
the right-hand side becomes

\begin{equation}\label{RHS}
RHS=\frac{1}{Z_{l}}\sum_{n=0}^{l}{l \choose n}<f|\psi_{n}^{l}><\psi_{n}^{l}|g>
    2(l+1)\int_{0}^{\infty}d|Z|\,|Z|^{2n+1}\frac{1}{(1+|Z|^{2})^{l+2}}P_{l}(|Z|^{2}).
\end{equation}

After the variable change $x=|Z|^{2}$ we must have:

\begin{equation}\label{P-intermed}
\int_{0}^{\infty}dx\,x^{n}\frac{1}{(1+x)^{l+2}}P_{l}(x)=e^{-\beta
    \varepsilon_{n}^{l}}\frac{1}{l+1}{l\choose n}^{-1}=e^{-\beta
    \varepsilon_{n}^{l}}\frac{1}{l+1}\frac{\Gamma(n+1)\Gamma(l+1-n)}{\Gamma(l+1)},
\end{equation}
where $P_{l}(x)$ is an unknown function. In order to determine it,
we extend the previous ansatz referring to the energy exponential
\cite{popov-PLA}.

It is obvious that the P-function also depends on the quantities
${\mathcal{A}}$ and ${\mathcal{B}}$, besides the variable $x$.
This leads to the idea that $P_{l}(x)\equiv
P_{l}(x,{\mathcal{A}},{\mathcal{B}})$ can also be expanded in a
power series similarly to the energy exponential (\ref{expand-E})
in the following manner:

\begin{equation}\label{P-l-expand}
P_{l}(x)=\sum_{k=0}^{\infty}\frac{{{\mathcal{B}}}^{k}}{k!}
\left[\left(\frac{d}{d{\mathcal{A}}}\right)^{k}
P_{l}(x,{\mathcal{A}},{\mathcal{B}})\right]_{{\mathcal{B}}=0}
\equiv \sum_{k=0}^{\infty}\frac{{{\mathcal{B}}}^{k}}{k!}
\left(\frac{d}{d{\mathcal{A}}}\right)^{2k}X_{l}(x, {\mathcal{A}}),
\end{equation}
where the function $X_{l}(x, {\mathcal{A}})$ is to be determined.

By inserting Eqs. (\ref{expand-E}) and (\ref{P-l-expand}) into the
equation (\ref{P-intermed}), performing the function change
obtain:

\begin{equation}\label{X-h}
X_l (x,{\mathcal A})= \frac{1}{\Gamma (l+2)} (1+x)^{l+2} h_l
(x,{\mathcal A})
\end{equation}
as well as by extending the natural values of $n$ to complex $s$
such as $n=s-1$, we get to the following Stieltjes moment problem:

\begin{equation}\label{Stieltjes-h}
\int_{0}^{\infty}
dx\, x^{s-1}h_{l}(x,{\mathcal{A}})=
e^{{\mathcal{A}}}\frac{1}{(e^{{\mathcal{A}}})^{s}}
\Gamma{(s)}\Gamma{(l+2-s)}.
\end{equation}

The solution of such a problem is \cite{math-sax}

\begin{equation}\label{h-final}
h_l (x,{\mathcal A})= e^{{\mathcal
A}}\Gamma{(l+2)}\frac{1}{(1+e^{{\mathcal A}} x )^{l+2}}.
\end{equation}

Finally, the P-function is

\begin{equation}\label{P-fin}
P_{l}(x)=\exp{\left[{\mathcal{B}}\left(\frac{d}{d{\mathcal{A}}}\right)^{2}\right]}
\left[e^{{\mathcal{A}}}\left(\frac{1+x}{1+e^{{\mathcal{A}}}x}\right)^{l+2}\right].
\end{equation}
where we have use the operator identity (\ref{op-ident}).

The scalar product of two Klauder-Perelomov coherent states for
the Morse oscillator is

\begin{equation}\label{scalar}
<Z, \alpha|Z',
    \alpha>=\frac{(1+\overline{Z}Z')^{l}}{(1+|Z|^{2})^{\frac{l}{2}}(1+|Z'|^{2})^{\frac{l}{2}}}.
\end{equation}

From the trace condition ($Tr\rho_{l}=1$) applied to the diagonal
expansion (\ref{rho-diagonal}) and the scalar product, it is not
difficult to prove that the P-function satisfies the normalization
condition:

\begin{equation}\label{trace-P}
\frac{1}{Z_{l}}\int d\mu(Z,\overline{Z})P_{l}(|Z|^{2})=1.
\end{equation}

Let us verify this equation using the our obtained expression for
the P-function (\ref{P-fin}). Thus, we have

\begin{eqnarray}\label{intermed-1}
I\equiv \frac{1}{Z_{l}}(l+1)\int_{0}^{2\pi}\frac{d\theta}{\pi}\frac{1}{2}
    \int_{0}^{\infty}dx\frac{1}{(1+x)^{2}}\exp{\left[{\mathcal{B}}\left(\frac{d}{d{\mathcal{A}}}\right)^{2}\right]}
\left[e^{{\mathcal{A}}}\left(\frac{1+x}{1+e^{{\mathcal{A}}}x}\right)^{l+2}\right]=\\
\nonumber
=\frac{1}{Z_{l}}(l+1)\exp{\left[{\mathcal{B}}\left(\frac{d}{d{\mathcal{A}}}\right)^{2}\right]}
\left[e^{{\mathcal{A}}}\int_{0}^{\infty}dx(1+x)^{l}(1+\exp{{\mathcal{A}}}x)^{-l-2}\right].
\end{eqnarray}

The last integral is of the following kind \cite{gradshteyn},
\cite{proudnikov}:

\begin{equation}\label{proud-300}
    \int_{0}^{\infty}\frac{(ax+b)^{\beta-1}}{(cx+d)^{\beta+1}}=\frac{(ad)^{\beta}-(bc)^{\beta}}{\beta(ad-bc)(cd)^{\beta}},\qquad
    \qquad
    [cd>0;\qquad ad\neq bc;\qquad {\mathrm{Re}}\beta \geq1].
\end{equation}

This lead to the results:

\begin{eqnarray}\label{I-inter}
 I=\frac{1}{Z_{l}}\exp{\left[{\mathcal{B}}\left(\frac{d}{d{\mathcal{A}}}\right)^{2}\right]}
    \left[\frac{1-(e^{\mathcal{A}})^{l+1}}
    {(1-e^{\mathcal{A}})(e^{\mathcal{A}})^{l}}\right]= \qquad \\
    \nonumber
    =\frac{1}{Z_{l}}\exp{\left[{\mathcal{B}}\left(\frac{d}{d{\mathcal{A}}}\right)^{2}\right]}
    \left[\sum_{n=0}^{l}(e^{-\mathcal{A}})^{n}\right]=1.
\end{eqnarray}

So, our obtained expression for the P-function (\ref{P-fin}) is
correct.

In this way, the diagonal representation of the normalized density
operator of the Morse oscillator in the Klauder-Perelomov coherent
states  representation is

\begin{equation}\label{rho-diag-final}
\rho_{l}=\frac{1}{Z_{l}}\int d\mu(Z,\overline{Z})\exp{\left[{\mathcal{B}}\left(\frac{d}{d{\mathcal{A}}}\right)^{2}\right]}
    \left[e^{\mathcal{A}}\left(\frac{1+|Z|^{2}}{1+e^{\mathcal{A}}|Z|^{2}}\right)\right]|Z,\alpha><Z,\alpha|.
\end{equation}

Then the thermal expectation value (the thermal average) of an
observable $A$ concerning the Morse oscillator is given by

\begin{equation}\label{thermal-A}
<A>_{l}=Tr(\rho_{l}A)=\frac{1}{Z_{l}}\int
    d\mu(Z,\overline{Z})P_{l}(|Z|^{2})<Z,\alpha|A|Z,\alpha>.
\end{equation}

If the operator $A$ is an integer power $s$ of the number operator
$N$, then, using Eqs. (\ref{thermal-A}), (\ref{N-s-diag},
(\ref{int-1}), as well as (\ref{expand-E}), we obtain the expected
result:

\begin{equation}\label{N-s-thermal}
<N^{s}>_{l}=\frac{1}{Z_{l}}\sum_{n=0}^{l}n^{s}e^{-\beta
    \varepsilon_{n}^{l}}=\frac{1}{Z_{l}}(-1)^{s}\left(\frac{\partial}{\partial
    \mathcal{A}}\right)^{s}Z_{l}.
\end{equation}

With this result we can define and calculate the thermal
second-order correlation function $\left(g^{(2)}\right)_{l}$ and
the thermal Mandel parameter $Q_{l}$, i.e. the thermal analogue of
the corresponding functions for the Klauder-Perelomov coherent
states $|Z,\alpha>$ (see, Eqs. (\ref{g-2-z}) and (\ref{Q-z})):

\begin{equation}\label{g-thermal}
\left(g^{(2)}\right)_{l}=\frac{<N^{2}>_{l}-<N>_{l}}{(<N>_{l})^{2}}=
    1+\frac{1}{\frac{\partial}{\partial
    \mathcal{A}}\ln{Z_{l}}}+\frac{\left(\frac{\partial}{\partial
    \mathcal{A}}\right)^{2}\ln{Z_{l}}}{\left(\frac{\partial}{\partial
    \mathcal{A}}\ln{Z_{l}}\right)^{2}},
\end{equation}

\begin{equation}\label{Q-thermal}
Q_{l}=<N>_{l}[\left(g^{(2)}\right)_{l}-1]=
    -1-\frac{\left(\frac{\partial}{\partial
    \mathcal{A}}\right)^{2}\ln{Z_{l}}}{\frac{\partial}{\partial
    \mathcal{A}}\ln{Z_{l}}}.
\end{equation}

The difference between these results and the similar results for
the Gazeau-Klauder quasi-coherent states of the Morse oscillator
\cite{popov-PLA} (i.e. the absence of the term $1/2$) is due
definition of the Morse energy eigenstates (\ref{E-n-l}),
comparing by
$E_{n}=\hbar\omega(n+1/2)-\frac{\hbar\omega}{2(l+1)(n+1/2)^{2}}$
(see, Eq. (3) in Ref. \cite{popov-PLA} ).

\section{\bf Discussion and outlook}
In conclusion, we have given a construction of Klauder-Perelomov
coherent states associated with the bound states of the Morse
potential.These states possess all properties of the usual
coherent states as continuity, overcompletion, non-orthogonality
and temporal stability, being defined on the entire complex space
of the variable $Z$.

In order to examine the statistical properties of these states, we
have concentrated our attention on  a quantum system which
consists on the quantum canonical ideal gas of Morse oscillators,
in thermodynamic equilibrium with the reservoir (thermostat).
Firstly, we have obtained the expressions of the diagonal elements
of the density operator in the Klauder-Perelomov coherent states
(the Husimi's Q-function), as well as the diagonal representation
of the density operator and the corresponding P-function. With
these functions we are able to express the thermal expectations
(thermal averages) for the physical quantities which characterize
the Morse oscillator quantum system. The second-order correlation
function and the Mandel parameter, as well as their corresponding
thermal pairs, offer the information on the character of the
coherent states, versus the Poisson distribution function.

Besides the implicit construction of the Klauder-Perelomov
coherent states for the Morse potential and examination of their
properties, the main results of this paper are the derivation of
the integration measure and the deduction of the expressions for
Q- and P-functions and, consequently, the thermal averages
calculated with these functions. In our opinion, the above
obtained results seems to be entire new, because, to our
knowledge, these results have not yet been published in the
literature.

By using the general expression for thermal averages one can easy
be deduced some interesting concrete thermal averages for the
Morse oscillators quantum canonical ideal gas as: free and
internal energy, entropy and molar heat capacity at the constant
volume and so on.

 {\vskip
1.0cm} {\bf Acknowledgements}:
M.D is grateful to Max Planck Institute for the Physics of Complex systems for its support .\\
\vfill\eject

\end{document}